\documentclass[conference,oneside]{IEEEtran}
\pdfoutput=1
\usepackage{algorithm}
\usepackage{comment}

\usepackage{amsfonts}
\usepackage{amsmath}
\usepackage{algorithm,algpseudocode}

\usepackage[space]{cite}
\let\oldbibliography\thebibliography
\renewcommand{\thebibliography}[1]{%
\oldbibliography{#1}%
\setlength{\itemsep}{2pt}
}

\ifCLASSINFOpdf
   \usepackage[pdftex]{graphicx}
   \graphicspath{{./figures/}}
   \DeclareGraphicsExtensions{.pdf,.jpeg,.jpg,.png}
\else
   \usepackage[dvips]{graphicx}
\fi

\usepackage{multirow}
\usepackage[font=footnotesize , labelfont=footnotesize, textfont=footnotesize]{caption}
\usepackage[font=footnotesize , labelfont=footnotesize, textfont=footnotesize]{subcaption}
\usepackage[squaren,Gray]{SIunits}
\usepackage{footnote}
\usepackage{algpseudocode}

\newcommand{\R}{\mathbb{R}}  
\newcommand{\vv}[2]{\mbox{VecVec}\left(#1,#2\right)}
\newcommand{\mv}[2]{\mbox{MatVec}\left(#1,#2\right)}
\newcommand{\Fig}[1]{Fig.~\ref{#1}} 

\usepackage[framemethod=default]{mdframed}

\usepackage{microtype}

\usepackage{flushend}

\usepackage{fancyhdr}
\begin{document}

\title{Doing Moore with Less -- Leapfrogging Moore's Law with Inexactness for 
Supercomputing \vspace{-0.6cm}}

\author{\IEEEauthorblockN{Sven Leyffer \IEEEauthorrefmark{1},
Stefan M.\ Wild\IEEEauthorrefmark{1},
Mike Fagan\IEEEauthorrefmark{2},
Marc Snir\IEEEauthorrefmark{3},
Krishna Palem\IEEEauthorrefmark{2},
Kazutomo Yoshii\IEEEauthorrefmark{1} and
Hal Finkel\IEEEauthorrefmark{4}}\smallskip

\IEEEauthorblockA{\IEEEauthorrefmark{1}Mathematics and Computer 
Science Division, 
    Argonne National Laboratory, Lemont, IL, USA}\smallskip
\IEEEauthorblockA{\IEEEauthorrefmark{2}Department of Computer Science, Rice University, Houston, TX, USA}\smallskip
\IEEEauthorblockA{\IEEEauthorrefmark{3}Department of Computer Science, 
University of Illinois at Urbana-Champaign,
Urbana, IL, USA}\smallskip
\IEEEauthorblockA{\IEEEauthorrefmark{4}Argonne Leadership Computing Facility (ALCF),
Argonne National Laboratory, Lemont, IL, USA}\smallskip
\IEEEauthorblockA{\small Corresponding author: Sven Leyffer, leyffer@anl.gov \vspace{0.3cm}}
}

\maketitle

\begin{abstract}
Energy and power consumption are major limitations to continued
scaling of computing systems. Inexactness, where the quality of the
solution can be traded for energy savings, has been proposed as an
approach to overcoming those limitations. In
the past, however, inexactness necessitated the need for highly
customized or specialized hardware. The current evolution of 
\emph{commercial off-the-shelf} ({\sc cots}) processors
facilitates the use of lower-precision arithmetic 
in ways that reduce energy consumption. We study these new
opportunities in this paper, using the example of an inexact Newton
algorithm for solving nonlinear equations.
Moreover, we have begun developing a set of techniques we call
\emph{reinvestment} that, paradoxically, use reduced precision to
improve the quality of the computed result: They do so by reinvesting
the energy saved by reduced precision.
\end{abstract}


\thispagestyle{fancy}


\IEEEPARstart{W}e consider the use of low-precision arithmetic in an inexact Newton method to investigate 
the potential power savings and reinvestment strategies.

The performance of leading supercomputers has increased by an order of
magnitude every 3--4 years in the past two decades. The main factors
contributing to this growth have been (i) 
the sustained exponential increase in the performance of processors, as
exemplified by Moore's Law; (ii) a growth in the size of the largest
supercomputers and of their energy consumption; and (iii) the use of
increasingly specialized and more energy-efficient components, such as
GPUs, long vector units, and multicores.

Two of these three factors are running out of steam. Moore's law is
is coming to an end as it has become
increasingly difficult, and hence expensive, to reduce the  energy
consumption of transistors. 
As devices approach atomic scales, further scaling will require a
different technology. There are few alternative
technologies that hold a promise of a better energy$\times$delay
product and none are close to commercial deployment.
Current top supercomputers require power in the range of 10--20MW; it
is hard to significantly increase these numbers, because of the cost
of energy and the limitation of 
existing installations. Increased specialization may help, but also
leads to increased development costs for the platforms and the
application codes using them.
Continued increase in effective supercomputer performance will increasingly
depend on a smarter use of existing systems, rather than on a brute
force increase in their physical performance.

\begin{mdframed}[style=exampledefault,frametitle={Significance Statement}]
Current supercomputers suffer from the high cost of energy
required to run them. Moreover, projected exascale machines
will see their costs multiplied by a factor of 1000. Clearly,
then, energy-efficient computing techniques are now, and will
be, extremely important.

Traditionally, users evaluate their computing results with a
\emph{quality} metric. For example, weather prediction
focuses heavily on accuracy of weather prediction.
Currently, any user seeking improved quality might
employ a higher class of supercomputer, bearing the associated
energy costs of the higher class of machine.

In contrast to this traditional approach, we introduce new
techniques that enable a higher quality answer \emph{without}
paying the increased energy costs of a higher-performing CPU.
In essence, smart use of a current machine has leapfrogged
a machine generation.
\end{mdframed}

Top500 \cite{TOP500}, the most commonly cited ranking of supercomputers,
measures supercomputer performance by using the number of floating-point
operations per second performed in course of solving an as-large-as-feasible,  
dense system of linear equations (Linpack). A common (and
correct) critique of this metric is that the benchmark used is not
representative of modern applications. A subtler critique is that the
notion of a problem having a unique solution is not representative of
modern practice, either. Many modern computational problems have an
implicit \emph{quality} metric, such as precision level. 
The right question is not how much time or energy it takes to
solve a problem; rather it is about the tradeoff between computation
and solution quality. Two specific questions naturally arise:
\begin{enumerate}
\item
What quality of answer can be achieved
with a given computation budget (for time or energy)? 
\item
How large a
budget is needed to achieve a certain quality threshold?
\end{enumerate}

Previous research has focused on the second
question, and has focused on the compute time budget. We focus on the
first question and focus on the energy budget. We believe this novel
focus best matches the reality of high-performance computing (HPC): Users 
attempt to achieve best
possible results with a given resource allocation; and the main resource constraint
in future systems is energy.
This proposed approach can raise
the \emph{effective performance} of systems by focusing attention on
the relevant tradeoffs.

To reiterate, ever since energy consumption has become a significant
barrier, there has been significant interest in innovative approaches
to help continued scaling. In our context, the approach that we build
on is the concept of inexact computing\cite{palem-1, palem-2, palem-3}
which established the principle of trading application quality for
disproportionately large energy savings. In this paper, we interpret
inexactness to be computational precision.While it is well known that
precision reduction can lead to savings, we take the extra step here
resulting in a two-phased approach. After applying the traditional
inexactness principle during the first phase, we now use a novel
second phase which involves reinvesting the saved energy from the
first phase, to establish that the quality of the application can be
improved significantly, when compared to the original ``exact’’
high-precision computation, all the while using a fixed energy budget.

The study of the precision level of numerical algorithms has been a
fundamental subject in numerical analysis since the inception of the
field \cite{Hammingbook,Wilkinsonbook}. Much attention has been devoted to the 
numerical errors induced in
simulations by the
discretization of continuous space and time; and to the tradeoff
between precision level and number of
iterations in iterative methods. Less attention has been paid in the
HPC world to the discretization of real 
numbers into 16-, 32-, or 64-bit floating-point values, since the use
of lower precision did not result in significant
performance gains on commodity processors.

The
evolution of technology is changing this balance: Vector operations
can perform twice as many single-precision operations as double-precision 
operations in the same time and using the same energy. 
Single-precision vector loads and stores can move twice as many words
than double precision in the same time and energy budget. The use of
shorter words also reduces cache misses.
Half precision has the potential to provide a further factor of two.
As communication becomes
the major source of energy consumption of microprocessors
\cite{bergman2008exascale,ang2014abstract}, the advantage of shorter
words will become more marked.  
This has led in recent years to a renewed interest in the use
of lower-precision arithmetic, where feasible, in HPC 
\cite{buttari2007mixed,baboulin2009accelerating,li2002design,rubio2013precimonious}. 

We use the following approach in this paper:
We pick as our base computational budget the amount of energy consumed
to solve a given problem to a given error bound using double-precision
arithmetic. We then examine how that same energy budget can be used to
improve the error bound, by using lower-precision arithmetic: We save
energy by replacing high precision with lower precision and
\emph{reinvest} these savings in order to improve
solution quality.

The organization of the rest of the paper is shown by the following ``mini''
table of contents:\\

{\small
\noindent
\begin{tabular}{p{20ex}@{\hspace{2ex}}p{40ex}}
 \raggedright Application: Solving Nonlinear Equations &
   This section
        elaborates the details of our canonical problem --- solving
        nonlinear equations. The method of choice for solving such problems
        is the inexact Newton's method.\\
 \\       
 \raggedright Experimental Results       &
This section describes our test machines and
        experimental methodology. We ran 2 classes of experiments:
           (1) Simple comparison experiments to determine energy costs of
                  various configurations; and (2) Reinvestment experiments to determine the effectiveness
                 of the reinvestment technique.\\
\\
\raggedright      The Mathematics of Energy Reinvestment &
         This section gives a mathematical
      account of how reinvestment works when viewed through the lens of convergence
      rates.\\
\\
\raggedright  Conclusions and Outlook &
          The final section distills our results into
      a set of conclusions, and points the way towards broader application of
      our techniques. \\
\end{tabular}

\section*{Application: Iteratively Solving Nonlinear Equations}

We investigate the effect of low-precision arithmetic on an inexact Newton 
method.
The inexact Newton method \cite{DES82} considered here, and  
Jacobian-free Newton-Krylov methods \cite{Knoll2004357} in general, are 
workhorses in modern nonlinear solvers and provide insight 
into how low-precision arithmetic can be exploited in scientific computing. Our goal is to solve the
nonlinear system of equations 
\begin{equation}
  F(x) = 0, \label{E:Newton}
\end{equation}
where $F: \R^n \to \R^n$ is a twice continuously differentiable function.

The basic inexact Newton method is described in Algorithm~\ref{alg:in1}.
It starts from an initial iterate $x^0$, and consists of outer and inner 
iterations. The outer
iterations correspond to approximate Newton steps that produce a 
sequence $\{x^k\}_k$ of iterates.
Given $x^k$, the inner iteration solves the Newton system
\begin{equation}
  \nabla F(x^k) s =-F(x^k)
  \label{eq:Newtonsystem}
\end{equation}
approximately for $s\in \R^n$; in our case, this is done by using BI-CGSTAB 
\cite{doi:10.1137/0913035}, but the specific form of inner solver is not a 
critical part of the present analysis. The inner iterations terminate on the 
accuracy of 
the putative direction $s$; that is, one stops when $s$ satisfies a relative 
residual criterion for \eqref{eq:Newtonsystem}, 
\begin{equation}
      \frac{\|\nabla F(x^k)s+F(x^k)\|}{\|F(x^k)\|} \leq \eta_k,  
\label{E:Inexact}
\end{equation}
where $0 \leq \eta_k < 1$ is a sequence of tolerances that is forced to zero 
as $k$ increases. The search direction, $s$, 
obtained in the inner iteration is used in a simple Armijo line search 
\cite{armijo1966minimization} 
in order to ensure global
convergence \cite{Kelleybook}. We note that this line search can fail if the
direction $s^i$ computed in the inner iteration is not a descent direction for
the residual norm, $\|F(x)\|$. In our experiments, we take this failure as an
indication that the precision level was too low, and switch to a higher level
of precision. More sophisticated approaches (e.g., based on iterative 
refinement) may
also be possible.

Our implementation is somewhat simplistic in the sense that we assume 
that there exists a solution $x^*$ with $F(x^*) = 0$ since we do not implement 
safeguards that allow
convergence to minimizers of the residual $F(x)$ if such a point does not 
exist. For a more
rigorous handling of such cases, see, e.g.,  \cite{doi:10.1137/S1052623403422637}.

\begin{algorithm}[tb]
\caption{Basic Inexact Newton Method. \label{alg:in1}}
\begin{algorithmic}
  \State   Input parameters  $\eta_0>0$, $\epsilon>0$; initialize $x^0\in \R^n$
  \State   Compute $F(x^0)$ and $\|F(x^0)\|$; set $k=0$
  \While {$\|F(x^k)\|>\epsilon$}
    \State \emph{Approx solve $\nabla F(x^k)s=-F(x^k)$ s.t. \eqref{E:Inexact} holds:}
    \State     $r^0 = -F(x^k)$; set $\|r^0\|=\|F(x^0)\|$
    \State     $q^0 = r^0$, $s^0=v^0=p^0=0$ 
    \State     $\rho_0=\alpha_0=\omega_0=1$, $i=0$
    \While{\label{line:beginloop} $\|r^i\|> \eta_k \|F(x^k)\|$}
      \State $i \gets i+1$
      \State       $\rho_i=\vv{q^0}{r^{i-1}}$
      \If{$\rho_i=0$}
         \State BI-CGSTAB method fails
      \EndIf
      \State       $\beta_i = \frac{\rho_i}{\rho_{i-1}} \frac{\alpha_{i-1}}{\omega_{i-1}}$
      \State       $p^i = r^{i-1} + \beta_i (p^{i-1}-\omega_{i-1}v^{i-1})$
      \State       $v^i = \mv{\nabla F(x^k)}{p^i}$
      \State       $\alpha_i = \frac{\rho_i}{\vv{q^0}{v^i}}$
      \State       $u^i = r^{i-1} - \alpha_i v^i$
      \If{$\| u^i \| = 0$}
         \State $s^i = s^{i-1} + \alpha_i p^i$
         \State            {\bf exit}
      \EndIf
      \State $t^i = \mv{\nabla F(x^k)}{u^i}$
      \State       $\omega_i = \frac{\vv{t^i}{u^i}}{\vv{t^i}{t^i}}$
      \State       $s^i = s^{i-1}+\alpha_i p^i + \omega_i u^i$
      \State       $r^i = u^i - \omega_i t^i$; compute $\|r^i\|$
    \EndWhile
    \State     $\delta = $ LineSearch$(x^k,s^i)$ along last $s^i$ from inner loop
    \State     Set $x^{k+1}=x^k+\delta s^i$ and compute $F(x^{k+1})$
    \State     Update 
$\eta_{k+1}=\min\left\{\|F(x^{k+1})\|^{1/2},\frac{1}{2}\right\}$
    \State     Set $k\gets k+1$ and iterate
  \EndWhile
\end{algorithmic}
\end{algorithm}

Our implementation is matrix-free, in that the Jacobian matrix $\nabla F(x)$ is 
never explicitly evaluated; the user needs only to implement vector products
with the Jacobian matrix. The function MatVec$(A,v)$ in Algorithm~\ref{alg:in1} 
implements the matrix-vector product $A v$. The function VecVec$(v,w)$ 
implements the
scalar product $v^Tw$. To illustrate typical benefits of such an approach, we 
exploit the fact that the Jacobian matrix in our examples is
tridiagonal, and thus we only store the nonzero entries in three vectors of 
size $n$.

We consider two sets of test problems of variable dimension.
The first problem, Laplace, is a well-conditioned linear system of equations, derived
from a central-difference discretization of the Poisson equation.
We note, however, that because we use an inexact Newton solver, the 
linear system is not solved in a 
single outer iteration. The second problem, Rosenbrock, is nonlinear 
and notoriously ill-conditioned,
and was chosen to provide a more strenuous test for our low-precision 
implementation.

\subsubsection*{Laplace}
The first system of equations is given by
\begin{eqnarray*}
F_1(x) &=&  b_1 + 4 x_1 - x_2
\\
F_i(x) &=&  b_i - x_{i-1} + 4 x_i - x_{i+1}, \qquad i=2,\ldots,n-1
\\
F_n(x) &=&  b_n - x_{n-1} + 4 x_n ,
 \label{eq:grad_laplace}
\end{eqnarray*}
where $b_1 = 1.0$, $b_i = -2.0$ for $i=2,\ldots,n-1$, and $b_n = 4.0$.

\subsubsection{Chained Rosenbrock}
The second, nonlinear system of equations is derived from the chained 
Rosenbrock function
\begin{equation}
\sum_{i=1}^{n-1} \left(a(1-x_i)^2 + 100\left(x_{i+1}-x_i^2\right)^2 
\right),
 \label{eq:f_rosen}
\end{equation}
whose first-order optimality conditions provide our system of equations and are 
given by 
\begin{eqnarray*}
F_1(x) &=&  2a(x_1-1) - 400x_i\left(x_{2}-x_1^2\right)
\\
F_i(x) &=&  200\left(x_i - x_{i-1}^2 \right) + 2a(x_i-1)
\\
       & & - 400x_i\left(x_{i+1}-x_i^2\right), \qquad i=2,\ldots,n-1 
\\
F_n(x) &=&  200\left(x_{n}-x_{n-1}^2\right).
 \label{eq:grad_rosen}
\end{eqnarray*}
The parameter $a>0$ controls the conditioning of the problem; in our tests 
we use its standard value $a=1$.

\section*{Experimental Results}
We now describe our experiments with reduced-precision variants of the inexact 
Newton method.

\subsection*{Experimental Testbed and Analysis Tools}
Our hardware testbed for this study was a Dell precision T1700
workstation operated by CentOS 7 and equipped with an Intel core i7 4770 
(3.40 GHz) and 16\,GB of DDR3 RAM. The processor had 4 cores, each with
2 hyperthreads, giving 8 logical CPUs.
The Intel core i7 4770 processor has 2 important architectural features:
\begin{enumerate}
  \item It implements the AVX2 instructions set
  \item \label{it:rapl} It supports the Running Average Power Limit (abbreviated RAPL)
    hardware counter~\cite{RAPL}
\end{enumerate}

Item~\ref{it:rapl} gives us a way to measure the energy consumption for a given program.
The specific RAPL tool that we used, \texttt{etrace2}, was written by one of the
authors of this paper.

In order to reduce the noise in our experiments, we launched the same process
on all logical CPUs, using Unix \texttt{taskset} to pin each instance of
the program being measured to a specific logical CPU. This prevents
process migration and avoids background activities on the idle cores.
We took the median of 31 separate measurements as our
measured energy consumption for a given data point.

The operating system was Linux, kernel 4.3.

We used the Intel compiler IFORT 16.0.3 20160415 to compile
our applications. The Intel compiler has exceptional automatic
vectorization capabilities.

\subsection*{Experimental Treatments -- Configuration Choices}
Our experiments were divided into two classes:
\begin{description}
\item[Gains.] This suite of experiments was devoted to finding out
  what gains were possible.
\item[Reinvestment.] \hspace{0.5cm} Once we had significant energy savings in hand, our
  next suite of experiments employed our \emph{reinvestment} technique
  to improve the quality of our computation.
\end{description}
For both classes of experiment, 
we employed the inexact Newton algorithm described in
Algorithm~\ref{alg:in1}, with the Laplace and Rosenbrock tests problems. We used 
a problem size of $n=100000$.
For each of Laplace and Rosenbrock, in our gains experiments we tested four variants:
\begin{enumerate}
    \item Scalar arithmetic, double precision
    \item Scalar arithmetic, single precision
    \item SIMD (vector) arithmetic, double precision
    \item SIMD (vector) arithmetic, single precision.
\end{enumerate}

For the reinvestment experiments, we limited our experimentation to
the SIMD (vector) configuration, since the vector codes are much more
efficient than the scalar ones.

\subsection*{Results of the Gains Experiments}

The energy expended by each of the four variants of the Laplace experiment are
shown in \Fig{fig:gains-l}. 
The energy expended by each of the four variants on Rosenbrock are
shown in \Fig{fig:gains-r}.

\begin{figure}[tb]
\centering\includegraphics[width=3in]{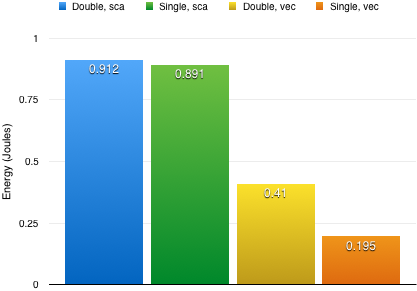}
\caption{Energy expended by Inexact Newton variants on Laplace; convergence to 
accuracy $\epsilon=10^{-6}$.}  
\label{fig:gains-l}
\end{figure}

\begin{figure}[hb]
\centering\includegraphics[width=3in]{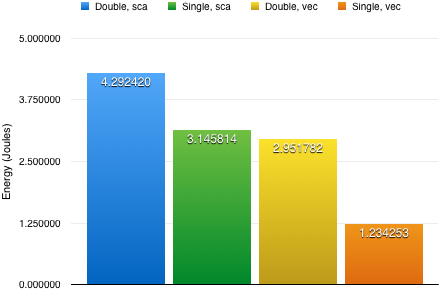}
\caption{Energy expended by Inexact Newton variants on 
Rosenbrock; convergence to 
accuracy $\epsilon=10^{-5}$.}\label{fig:gains-r}
\end{figure}

\subsection*{Explaining the Gains}
The use of 32-bit (single precision) scalars, rather than 64-bit (double precision) scalars, had
limited impact on the energy consumption of arithmetic operations or
loads. This is not surprising, since all the data paths, registers and
ALUs handle 64-bit values. The use of 32-bit values, has an additional benefit
on cache behavior. In general, the beneficial cache effects will depend on the memory access patterns,
and how much of the data fits in cache. For the problems we studied in this paper, the data exhibited
strong spatial locality --- effectively doubling the capacity of caches. This doubling effect
greatly reduced the cache misses.

Also, note that vector operations are more energy-efficient, per operand, than scalar
operations. Furthermore, a vector arithmetic operation, or a vector
load handles twice as many floats as doubles in the same time and
energy consumption. Hence, we expect some improvement when moving from
vector doubles to vector singles, due to better cache hit ratio
significant improvement when going to vector operations, and at least
of factor of two improvement when shifting from double vectors to float
vectors. The results in Figs.~\ref{fig:gains-l} and \ref{fig:gains-r} 
are consistent with these expectations. 

\subsection*{Improving Application Quality Through Reinvestment}
Our initial gain experiments show that the SIMD vector codes have better
energy savings. Consequently, we now focus on the vectorized variants for
our reinvestment experiments.

We recall our reinvestment strategy:
\begin{itemize}
\item
We run the algorithm in double precision to a given error bound and
measure energy consumption. This is our energy budget.
\item
We next run the algorithm in single precision for a 
number of
iterations, followed by double precision for a number of iterations,
consuming the same energy as before, and measure the error bound. The ratio
between the first error bound and the second is the achieved
\emph{improvement factor}.
\end{itemize}

We have not yet developed an algorithm to decide automatically when to
switch from single
precision to double precision. Therefore, we experiment with different
switching points, in order to estimate the gains that reinvestment
can achieve.

Our reinvestment experiments have the following steps:

\begin{enumerate}
    \item\label{i:sing-tol} Choose a small accuracy tolerance for the 
single-precision variant. 
      This tolerance is denoted by $\epsilon$ in Algorithm~\ref{alg:in1}.\\
      \textbf{Note:} A chosen tolerance can result in a line search error, 
      which means that the solution $s^i$ of the inner iteration is not a 
      descent direction for the residual norm. We take this failure as an 
      indication that the chosen tolerance is too small.
    \item\label{i:dbl-tol} Run the double-precision variant using the same 
tolerance as step~\ref{i:sing-tol}.
    \item\label{i:pick-improve} Pick an improvement factor $f$ to try experimentally. 
    \item\label{i:reinvest} Run the reinvestment algorithm in single precision
      until it achieves the same accuracy tolerance as in 
step~\ref{i:sing-tol}; then continue in
      double precision until this tolerance is further reduced by a
      factor of $f$.
  If the energy from this step does \emph{not} exceed the energy
       from step~\ref{i:dbl-tol}, then the reinvestment experiment was successful!\\
      \textbf{Note:(Again)} As in step~\ref{i:sing-tol}, if a line search error is present
       in the double-precision reinvestment stage, then the chosen improvement factor is too large.
\end{enumerate}


\subsubsection*{Laplace}
The Laplace example, being linear, quickly converges to a very good approximation of the
answer. In our experiments, accuracy tolerance values smaller than $10^{-6}$ 
resulted in line search errors. These failures are consistent with the fact that $10^{-6}$ is 
close to single-precision machine precision, so it would be unrealistic to expect better. 
For Laplace, we were able to achieve an improvement factor of $10^4$.
The limitation on the improvement factor was the low energy budget.
An improvement factor of $10^5$ is computationally possible, but the reinvestment energy
budget is exceeded for this improvement factor. Improvement factors of $10^6$ (or higher)
led to line search errors.
Details of the experiment appear in 
Table~\ref{tbl:reinv-l}; a graphical representation appears in 
\Fig{fig:reinv-l}.

\begin{table}
   \caption{Reinvestment for Laplace}\label{tbl:reinv-l}
   \centering
     \footnotesize
   \resizebox{\linewidth}{!}{%
   \begin{tabular}{|l|c|c|c|c|}
   \hline
   \textbf{Class} & \textbf{Energy} & \textbf{Std} & \textbf{Iterations Single} & 
\textbf{Iterations Double}\\
         & \textbf{(Joules)}       & \textbf{Deviation} & \textbf{Outer/Inner}       & \textbf{Outer/Inner}\\
   \hline \hline
   Reinvest & 0.843407 & 0.07 & 9/12 & 2/9\\
   \hline
   Ceiling (double) & 1.1026 &0.06 & NA & 8/10\\
   \hline
   Base (single) & 0.510376 & 0.05 & 9/12 & NA\\
   \hline
   \end{tabular}}
\end{table}

\begin{figure}[htb]
\centering\includegraphics[width=3in]{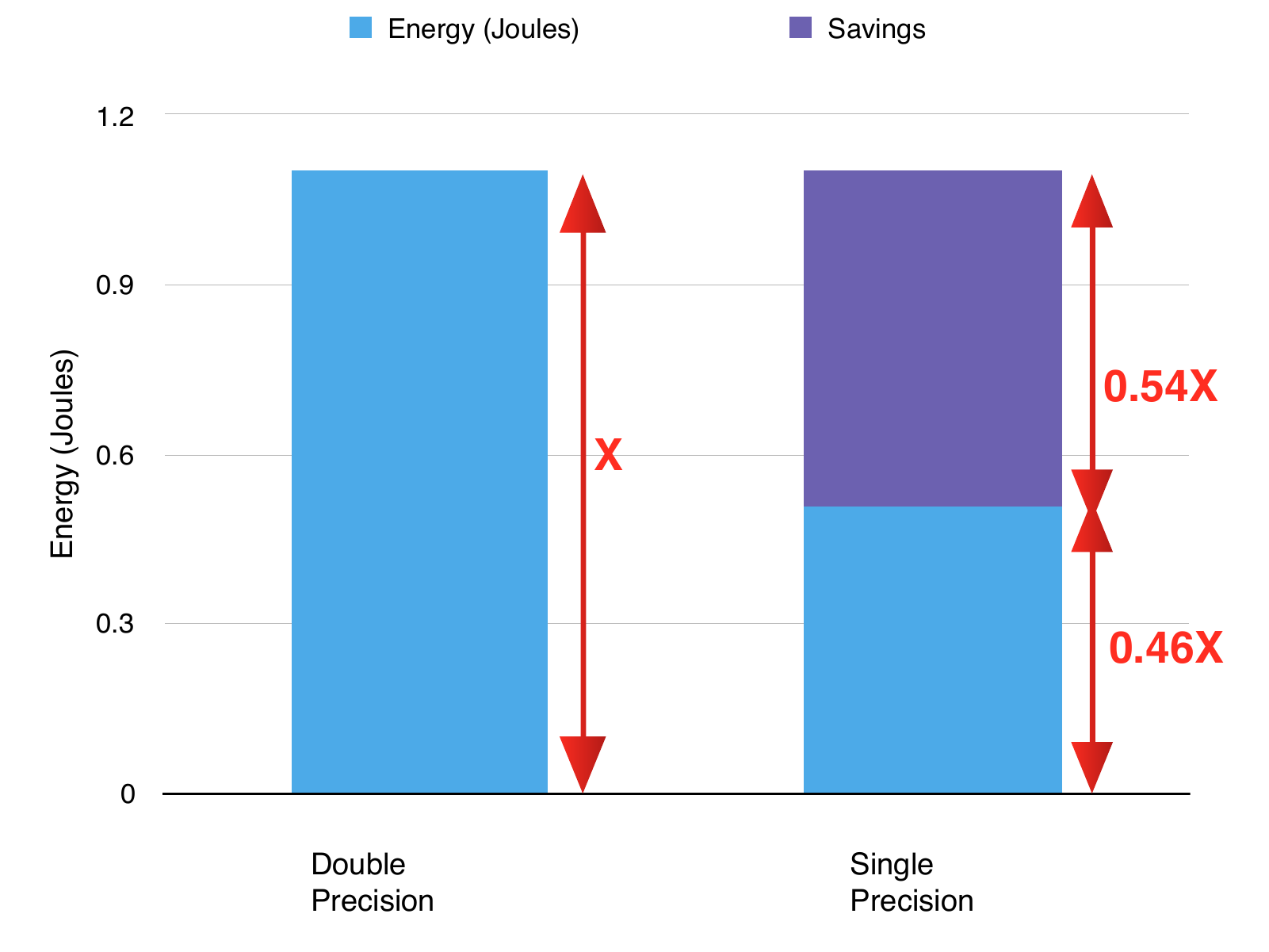}
\caption{Reinvestment for Laplace; original  
  accuracy $\epsilon=10^{-6}$, improvement factor$=10^4$. The ``X'' graphic shows
  the total energy budget. The arrows labeled with fractions of ``X'' show base convergence
energy, and energy available for reinvestment.}\label{fig:reinv-l}
\end{figure}

\subsubsection*{Rosenbrock}
The Rosenbrock example had more scope for experimentation. For our first
reinvestment experiment with Rosenbrock, we chose a conservative 
accuracy tolerance of
$10^{-2}$. For this tolerance, both pure single and pure double required
29 (68) outer (inner) iterations. Next, we experimentally determined that
$10^{11}$ was the highest improvement factor we could obtain without line
search errors in the reinvestment phase. Even with this large improvement
factor, we were unable to use all of the saved energy.
Table~\ref{tbl:reinv-r-e11} shows the measured energy and iteration counts;
\Fig{fig:reinv-r-e11-gain} shows the relative gain savings.
Figure~\ref{fig:reinv-r-e11-reinv} shows the fraction of the \emph{saved} energy
that contributed to the reinvestment improvement, as well as the remaining energy.

\begin{table}
   \caption{Reinvestment for Rosenbrock, $10^{-2}$ initial convergence, 
     $10^{11}$ improvement.}\label{tbl:reinv-r-e11}
   \centering
   \resizebox{\linewidth}{!}{%
   \begin{tabular}{|l|c|c|c|c|}
   \hline
   \textbf{Class} & \textbf{Energy} & \textbf{Std} & \textbf{Iterations Single} & 
\textbf{Iterations Double}\\
         & \textbf{(Joules)}       & \textbf{Deviation} & \textbf{Outer/Inner}  & \textbf{Outer/Inner}\\
   \hline \hline
   Reinvest & 4.20105 & 0.14 & 29/68 & 4/18\\
   \hline
   Ceiling (double) & 6.92410 & 0.16 & NA & 29/68\\
   \hline
   Base (single) & 2.69243 & 0.14 & 29/68 & NA \\
   \hline
   \end{tabular}}
\end{table}

\begin{figure}[htb]
\centering\includegraphics[width=3in]{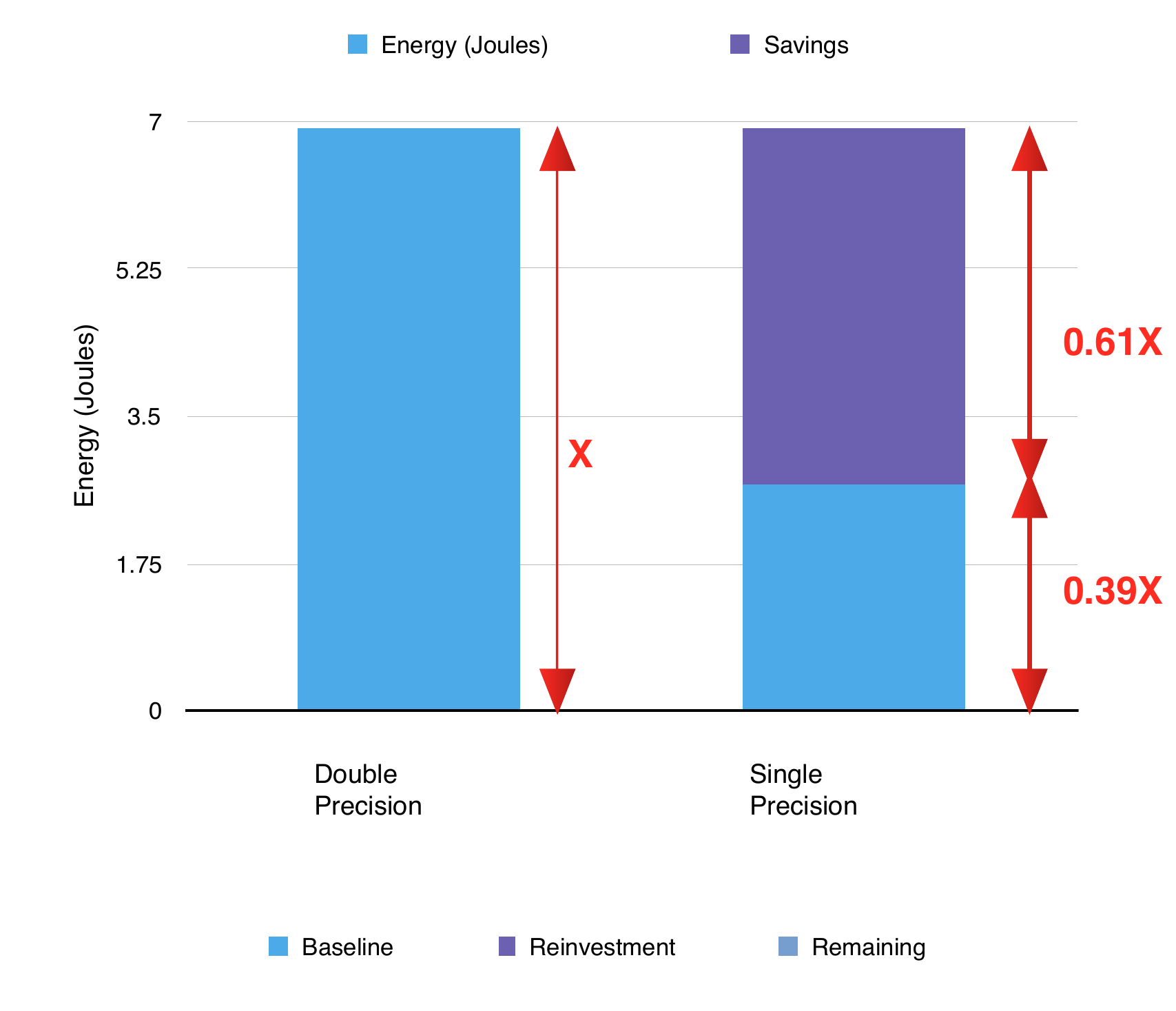}
\caption{Reinvestment for Rosenbrock; original 
accuracy $\epsilon=10^{-2}$. The ``X'' graphic shows
  the total energy budget. The arrows labeled with fractions of ``X'' show base convergence
energy, and energy available for reinvestment.}\label{fig:reinv-r-e11-gain}
\end{figure}

\begin{figure}[htb]
\centering\includegraphics[width=3in]{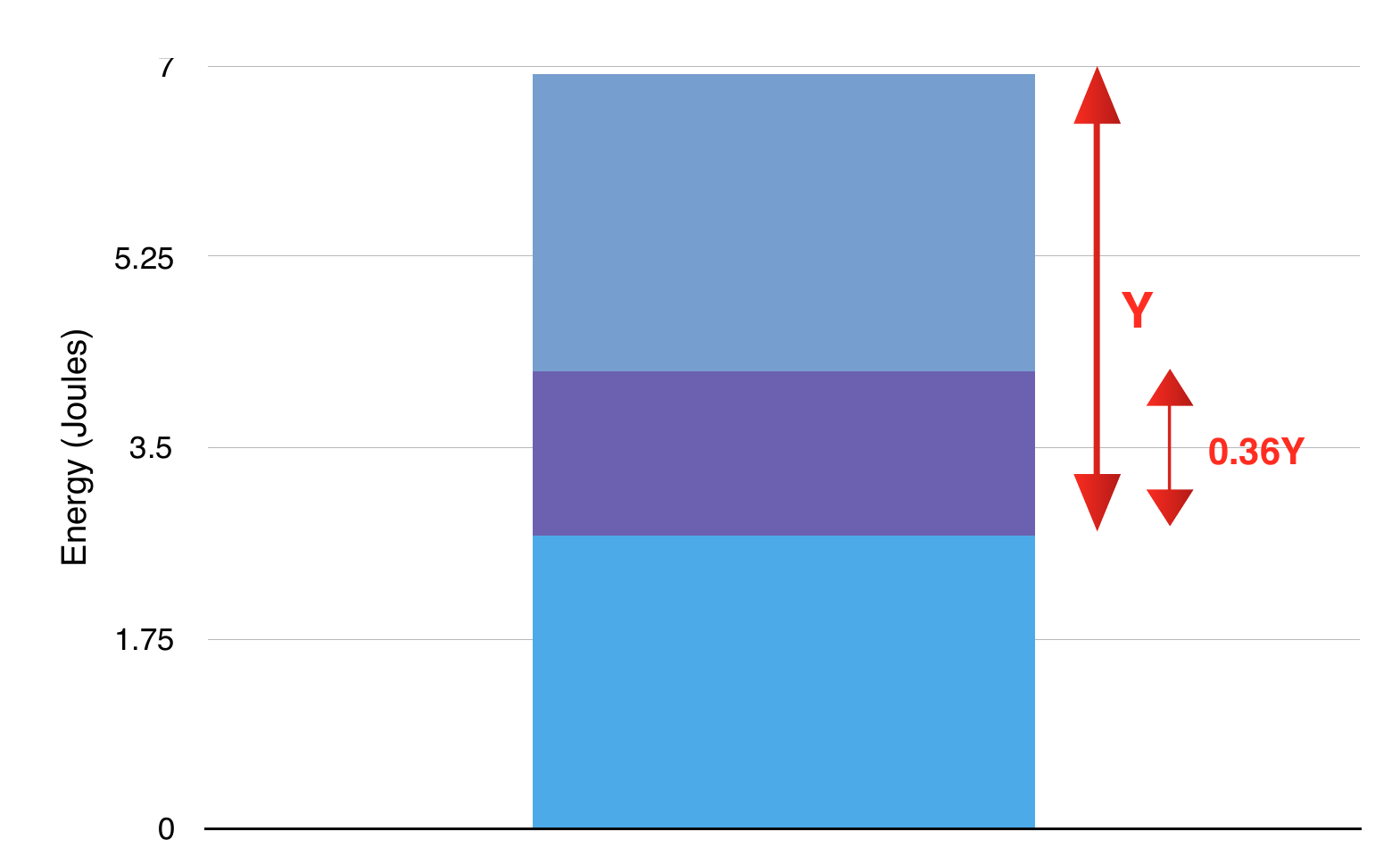}
\caption{Reinvestment for Rosenbrock; original 
  accuracy $\epsilon=10^{-2}$; $10^{11}$ improvement factor.
The ``Y'' arrow shows the total energy available for reinvestment.
   The fractional ``Y'' shows the fraction that could actually be used.}\label{fig:reinv-r-e11-reinv}
\end{figure}

The Rosenbrock example supported a convergence tolerance as small as $10^{-5}$ when
using single precision. Smaller tolerances again led to line search errors. In 
contrast to the Laplace
example, convergence on Rosenbrock was slower -- 31 outer iterations were required
to achieve the desired quality. The large number of iterations, however, meant
that there could be substantial savings for reinvestment. We experimentally
determined the highest improvement factor to be $10^8$; higher factors
resulted in line search errors. Even with the high improvement factor, there
was still reinvestment energy remaining. Details of the Rosenbrock reinvestment
experiment appear in Table~\ref{tbl:reinv-r}. A single graph showing the
baseline, reinvestment, and extra energy can be seen in 
\Fig{fig:reinv-r}.

\begin{table}
  \caption{Reinvestment for Rosenbrock, $10^{-5}$ initial convergence, $10^8$ improvement.}\label{tbl:reinv-r}
   \centering
   \resizebox{\linewidth}{!}{%
   \begin{tabular}{|l|c|c|c|c|}
   \hline
   \textbf{Class} & \textbf{Energy} & \textbf{Std} & \textbf{Iterations Single} & 
\textbf{Iterations Double} \\
         & \textbf{(Joules)}    & \textbf{Deviation} & \textbf{Outer/Inner}   & \textbf{Outer/Inner}\\
   \hline \hline 
   Reinvest & 4.20730 & 0.23 & 31/76 & 3/14\\
   \hline
   Ceiling (dbl) & 7.84135 & 0.22 & NA & 31/76\\
   \hline
   Base (single) & 3.02834 & 0.31 & 31/76 & NA\\
   \hline
   \end{tabular}}
\end{table}

\begin{figure}[htb]
\centering\includegraphics[width=3in]{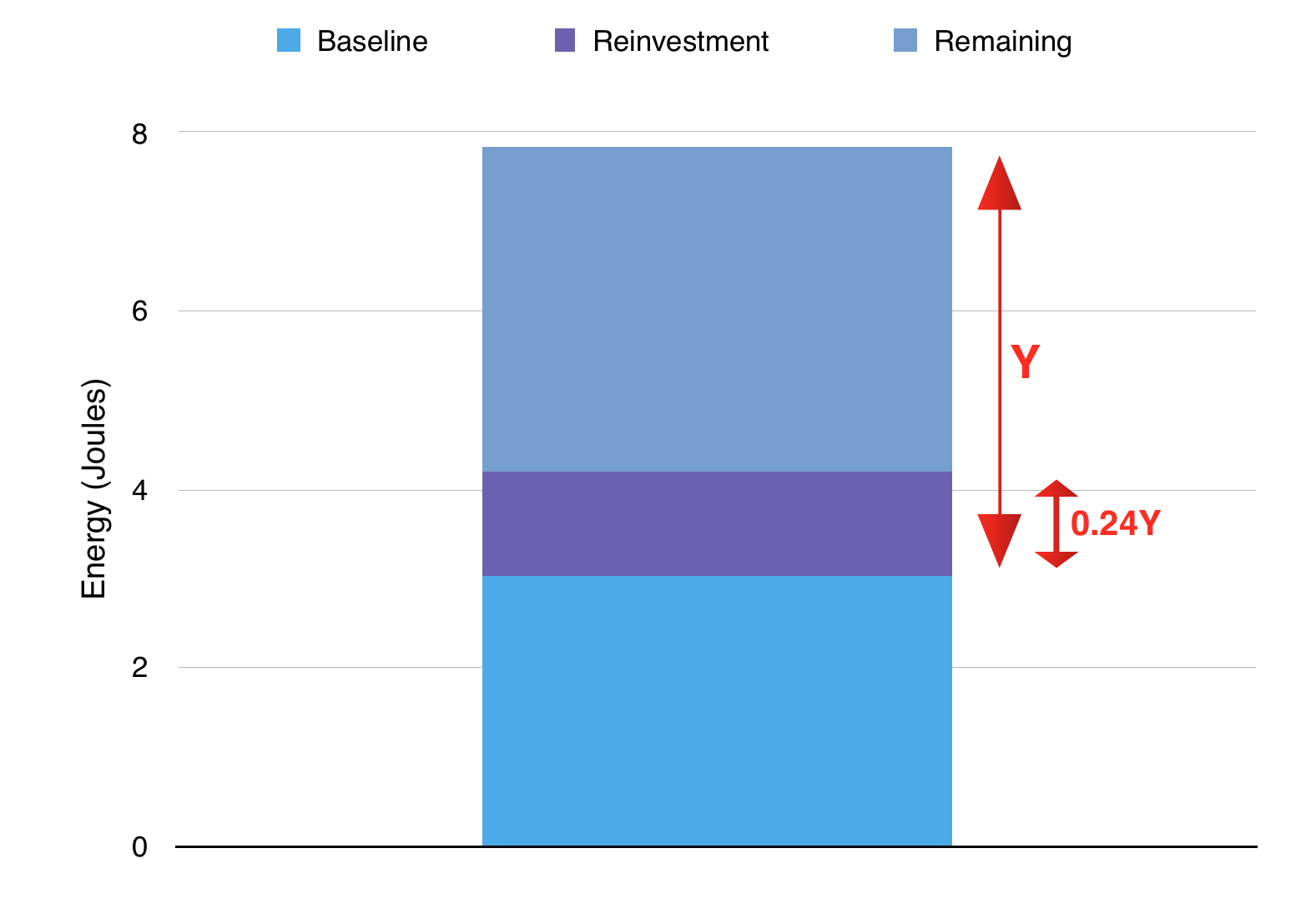}
\caption{Reinvestment for Rosenbrock; original
  accuracy $\epsilon=10^{-5}$; improvement factor$=10^8$.
The ``Y'' arrow shows the total energy available for reinvestment.
   The fractional ``Y'' shows the fraction that could actually be used.}\label{fig:reinv-r}
\end{figure}

\subsection*{Energy as a Function of Improvement Factor}
In the previous section, we were usually working under the assumption that
users would seek the largest improvement factor. In reality, a user
may seek \emph{some} improvement, but not the maximum possible improvement.
In this section, we show the energy costs of some intermediate improvement
factors.
\subsubsection*{Energy/Improvement Gradations for Laplace}
Since the maximum improvement factor for Laplace was $10^4$, we looked at
improvement factors $10^1$, $10^2$, $10^3$, and $10^4$. What we see
is that the number of Outer / Inner iterations is the same for each of the
improvement factors. This behavior is due to the fact that a single
additional outer iteration requires 4 inner iterations, and achieves an
improvement factor of $10^4$. For Laplace, a user might just as well seek the 
maximum
improvement factor of $10^4$.

The table for each of the factors is show in Table~\ref{tbl:graded-l}.
Figure~\ref{fig:graded-l} shows the same information graphically.
Note that the energy numbers are not identical due to measurement error,
but lie within the error bounds implied by the standard deviation.

\begin{table}[hb]
  \caption{Energy cost as a function of improvement factor for Laplace;
    original $\epsilon=10^{-6}$}\label{tbl:graded-l}
   \centering
   \resizebox{\linewidth}{!}{%
   \begin{tabular}{|c|c|c|c|}
   \hline
   Improvement & Energy (Joules) & Std & Iterations\\
   Factor (exponent) &  & Deviation & Outer/Inner\\
   \hline
    0 (base) & 0.510376 & 0.05 & NA/NA\\
   \hline
    1 & 0.832625 & 0.061  & 1/4\\
   \hline
    2 & 0.845 & 0.053 & 1/4\\
   \hline
    3 & 0.843375 & 0.056 & 1/4\\
    \hline
    4 & 0.843407 & 0.07  & 1/4\\
    \hline
   \end{tabular}}
\end{table}

\begin{figure}[tb]
\centering\includegraphics[width=3in]{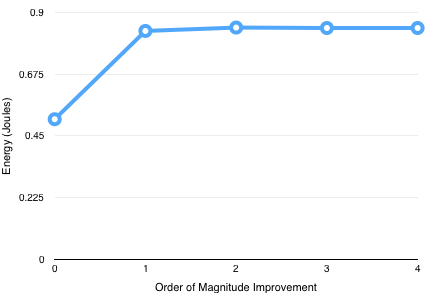}
\caption{Energy as a function of improvement factor for Laplace; original 
convergence to 
accuracy $\epsilon=10^{-6}$.}  
\label{fig:graded-l}
\end{figure}

\subsubsection*{Energy/Improvement Gradations for Rosenbrock}
The Rosenbrock results show 3 clusters of iteration counts:
factors 1--2, factors 3--6, and factors 7--8.
As with Laplace, these groups correspond to 1, 2, and 3 outer iterations that
achieve improvement factors of $10^2$, $10^6$, and $10^8$, respectively. Again,
the energy of each group is statistically identical. 
The details are shown in Table~\ref{tbl:graded-r}, with the graphical
representation in \Fig{fig:graded-r}.

\begin{table}[hb]
  \caption{Energy cost as a function of improvement factor for Rosenbrock;
    original $\epsilon=10^{-5}$}\label{tbl:graded-r}
   \centering
   \resizebox{\linewidth}{!}{%
   \begin{tabular}{|c|c|c|c|}
   \hline
   Improvement & Energy (Joules) & Std & Iterations\\
   Factor (exponent) &  & Deviation & Outer/Inner\\
   \hline
    0 (base) & 3.0283 & 0.31  & NA/NA\\
   \hline
    1 & 3.3562 & 0.158 & 1/4\\
   \hline
    2 & 3.385 & 0.052 & 1/4\\
   \hline
    3 & 3.7488 & 0.066 & 2/8\\
    \hline
    4 & 3.7762 & 0.038 & 2/8\\
    \hline
    5 & 3.7775 & 0.077 & 2/8\\
    \hline
    6 & 3.7762 & 0.054 & 2/8\\
    \hline
    7 & 4.2575 & 0.075 & 3/14\\
    \hline
    8 & 4.2073 & 0.23 & 3/14\\
    \hline
   \end{tabular}}
\end{table}

\begin{figure}[tb]
\centering\includegraphics[width=3in]{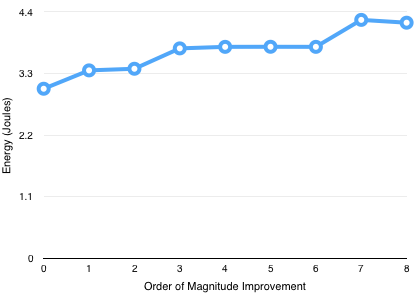}
\caption{Energy as a function of improvement factor for Rosenbrock; original 
convergence to 
accuracy $\epsilon=10^{-5}$.}  
\label{fig:graded-r}
\end{figure}

\section*{The Mathematics of Energy Reinvestment}
We present a model of the reinvestment of the energy saved using low-precision arithmetic,
and show that under fairly mild assumptions, we can expect to obtain a more accurate solution
at a reduced cost, and we quantify the potential savings in the example of our inexact Newton
applications.


Given its iterative nature and adaptive accuracy requirements, the inexact 
Newton method lends itself almost ideally to an approach that seeks to take 
advantage
of inexact and adaptive-precision arithmetic. 
We are motivated by the observation that most of the work of Newton's method 
typically 
occurs before the transition to fast quadratic convergence. 

In Algorithm~\ref{alg:meta}, we propose a simple 
meta-strategy that starts by running Algorithm~\ref{alg:in1}
with low-precision arithmetic, and switches to increasingly higher accuracy 
levels as we approach the solution. 
It 
consists of running the inexact Newton method at prescribed precision levels, 
with an accuracy tolerance, discussed further below, based on the 
current precision level. Many other approaches are possible, but this simple 
meta-strategy facilitates analysis and is representative of other 
inexactness-switching-based strategies for linear and nonlinear solvers
\cite{Hovland2001503,GY1999}.


\begin{algorithm}[tb]
\caption{Meta Strategy for Inexact Newton. \label{alg:meta}}
\begin{algorithmic}
  \State  Given precision levels $p_1 < p_2  < 
\ldots < p_L$. 
  \State  Initialize $x^0\in \R^n$.
  \For{$l=1,\ldots,L$}
    \State Obtain solution $x^{l}$ by running
 Algorithm~\ref{alg:in1} at precision level $p_l$ starting from $x^{l-1}$
(to an accuracy level $\epsilon = $Ep$(p_l,x^{l-1})$ depending on the precision 
level).
  \EndFor
\end{algorithmic}
\end{algorithm}



\subsection*{Energy Model for Floating-Point Operations}
\label{sec:energy}

We generalize the energy model in~\cite{palem-date} and derive an energy model for each individual 
iteration of inexact Newton that also takes cache misses into account.
We note that for both equations (Laplace and Rosenbrock) solved by our inexact Newton code
have tridiagonal Jacobians. Hence, the MatVec operation for our particular test cases
is $O(n)$, rather than $O(n^2)$ in the dense case.
All other operations in any given iteration are also $O(n)$.

We let $k n$ denote the number of floating-point operations at an 
outer iteration, and $l n$ denote 
the number of floating-point storage locations. In our examples, both $k$ and 
$l$ are small integers when 
compared with the problem dimension $n$. We let $E_c(p)$ and $E_t(p)$ 
denote the energy required
for a single floating-point compute and transfer, respectively, at precision level p. 
We neglect the cost of L1 cache transfers, and instead only consider L2 and L3 cache transfers. 
We let $r(p)$ denote the cache-miss rate at precision level $p$ (i.e., number 
of bits).

We assume that energy and cache-miss rates scale linearly with the 
precision-level $p$. 
For the sake of convenience, we assume that energy and miss-rates can be simply 
expressed as
\begin{equation}
 E_c(p) = p E_c, \quad E_t(p) = p E_t, \quad r(p) = p r. 
 \label{eq:linear}
\end{equation}
Under these assumptions, it follows that the energy required for a 
single outer iteration is 
\begin{equation}
E(p) = k n E_c p + l n r p E_t p = k n E_c p + l n r E_t p^2, 
\label{eq:energy}
\end{equation}
where the second term depends quadratically on the precision, because both the 
cache-miss rate, $r(p)$,
and the transfer energy, $E_t(p)$, are linear functions of the precision level. 
Although this may indicate
that we can expect superlinear energy savings, the rate constant is typically too small (in practice,
$r$ takes values $r \in [10^{-3}, 10^{-6}]$), and the linear term dominates.

%

\subsection*{Model for Reinvestment Strategy}

Here, we develop a mathematical model for the reinvestment strategy
described in Algorithm~\ref{alg:meta}. Formally, we compare the
accuracy of a computation done using double precision to the accuracy
of a computation using the same amount of energy, but using single
precision, where feasible. The analysis can be extended to the case of
more than two levels of precision.


We denote by $\epsilon_k = \| F(x^k) \|$ the error at the $k$th 
outer iteration and assume, without loss of generality,  that
$\epsilon_0 \leq 1$.


We also assume that the method can use a precision of $p$ as
long as $\epsilon > 2^{-p+s}$. That is, that the inner and outer 
loops will terminate provided the current solution error is large
with respect to the precision of floating-point arithmetic.
The $s$ term relates to the length of the exponent, and to the amount of 
rounding errors accumulated during an outer iteration. 

To simplify the discussion, we shall assume that $s = 3\lg p -7+\delta$.
The value of 
$2^{-p+3\lg p-7}$ is $2^{-11}$, $2^{-24}$, and $2^{-53}$ for $p=16,32,64$, 
respectively, and approximates machine epsilon in the IEEE 754 standard (see, 
e.g., \cite{Demmel:1997}). The $2^\delta$ term captures the conditioning of the 
problem; for simplicity, we use $\delta=8$ here, so that we can use single 
precision if
$\epsilon >2^{-16} \approx 10^{-5}$ and double precision if $\epsilon >2^{-45} 
\approx 3\cdot 10^{-14}$.


The advantage of lower precision will depend on the rate of
convergence of the iterative method.

\subsubsection*{Models Assuming Linear Convergence}
We  consider first the case where the iterative method
converges linearly: $\epsilon_{k+1} \leq \epsilon_k/\lambda$, with
$\lambda > 1$. Since $\epsilon_0 \leq 1$, the error after $k$ iterations 
is bounded by $\epsilon_k \leq \lambda^{-k}$, provided that the precision 
level $p$ satisfies $\epsilon_k \geq 2^{-p+s}$. Combining these two bounds, we 
see that at most $\frac{p-s}{\lg \lambda}$ such iterations are possible.

Now suppose that we have two precision levels, $p_1<p_2$ and an accuracy level 
$\epsilon$ that is attainable by the higher-precision level $p_2$ (i.e., 
$\epsilon \geq 2^{-p_2+s_2}$). The energy consumed by this baseline precision 
to be guaranteed an accuracy level $\epsilon$ is 
\begin{equation}
E(p_2)\frac{\lg \frac{1}{\epsilon}}{\lg \lambda}.
\label{eq:highp}
\end{equation}

Now consider a hybrid procedure that performs $k_1$ iterations at 
precision level $p_1$ followed by $k_2$ iterations at precision level $p_2$. 
The energy consumed by such a hybrid procedure is
\begin{equation}
E(p_2)\left(k_1 \frac{E(p_1)}{E(p_2)} + k_2\right).
\label{eq:hybridp}
\end{equation}
If we require that the energy consumed by the hybrid procedure 
is no more than that for the baseline ($p_2$) procedure, then \eqref{eq:highp} 
and \eqref{eq:hybridp} imply that we must have the following bound on the 
number of iterations at the higher-precision level:
\begin{equation}
k_2 \leq  \frac{\lg \frac{1}{\epsilon}}{\lg \lambda} - k_1 
\frac{E(p_1)}{E(p_2)}.
\label{eq:hybridp-ubound}
\end{equation}
Applying \eqref{eq:hybridp-ubound}, we have that the accuracy of the 
hybrid procedure is therefore bounded by 
\begin{equation}
\epsilon_{k_1+k_2} 
\leq  \lambda^{-k_1-k_2} 
\leq \lambda^{-k_1\left(1- \frac{E(p_1)}{E(p_2)}\right) - \frac{\lg 
\frac{1}{\epsilon}}{\lg \lambda}}.
\label{eq:hybrid-acc}
\end{equation}
Since $\lambda>1$ and $E(p_1)<E(p_2)$, the bound in \eqref{eq:hybrid-acc} is 
decreasing in the number of lower-precision iterations, $k_1$, and thus  one 
should choose $k_1$ to be as large as possible.
The bound in \eqref{eq:hybrid-acc}, however, only applies if the accuracy 
level $\lambda^{-k_1}$ is attainable at the lower-precision level; that is, if 
the lower-precision number of iterations satisfies $k_1 \leq \frac{p_1-s_1}{\lg 
\lambda}$. Similarly, the accuracy 
level $\epsilon_{k_1+k_2}$ is only attainable at the higher-precision level if 
 $\epsilon_{k_1+k_2}\geq 2^{-p_2+s_2}$. Applying these two inequalities to the 
bound in \eqref{eq:hybrid-acc}, we have that
\begin{equation}
\begin{array}{rl}
\epsilon_{k_1+k_2}
&\leq 
\max\left\{ 2^{-p_2+s_2},
\lambda^{-\frac{p_1-s_1}{\lg \lambda} \left(1- \frac{E(p_1)}{E(p_2)}\right) - 
\frac{\lg 
\frac{1}{\epsilon}}{\lg \lambda}}
\right\} \\
&=
\max\left\{ 2^{-p_2+s_2},
2^{-(p_1-s_1) \left(1- \frac{E(p_1)}{E(p_2)}\right) - 
\lg 
\frac{1}{\epsilon}}
\right\} \\
&=
2^{-\min \left\{p_2-s_2,
(p_1-s_1) \left(1- \frac{E(p_1)}{E(p_2)}\right) + 
\lg 
\frac{1}{\epsilon}
\right\} },
\end{array}
\label{eq:hybrid-acc2}
\end{equation}
where the first equality follows from the relation $\lambda^{\frac{b}{\lg 
\lambda}}=2^b$. 

The improvement factor thus satisfies
\begin{equation}
\frac{\epsilon}{\epsilon_{k_1+k_2}} 
\geq  
2^{ \min \left\{\lg \epsilon +p_2-s_2,
(p_1-s_1) \left(1- \frac{E(p_1)}{E(p_2)}\right)\right\} }.
\label{eq:impfact}
\end{equation}
We note that this bound relates the improvement factor to the original 
accuracy ($\epsilon$), the precision levels ($p_i-s_i$), and the energy ratio 
($\frac{E(p_1)}{E(p_2)}$). Figure~\ref{fig:impfactor-linear} illustrates this 
bound for a variety of precision levels under the assumptions $s_i=3\lg p_i+1$ 
and 
$\frac{E(p_1)}{E(p_2)}=\frac{p_1}{p_2}$.

We also note that these bounds are not necessarily tight, and thus even larger 
improvement factors can be seen in practice. As a specific example, for 
the hybrid Rosenbrock measurements in Table~\ref{tbl:reinv-r} with 
$\frac{E(p_1)}{E(p_2)} \approx 0.3$,  then $p_1=32$, $s_1=s_2=0$, and 
\eqref{eq:impfact} give the bound 
$\frac{\epsilon}{\epsilon_{k_1+k_2}} \geq 5.5 \cdot 10^6$. This bound is still 
more than an order of magnitude short of the measured improvement factor $10^8$.

\begin{figure}[tb]
\centering\includegraphics[width=\linewidth]{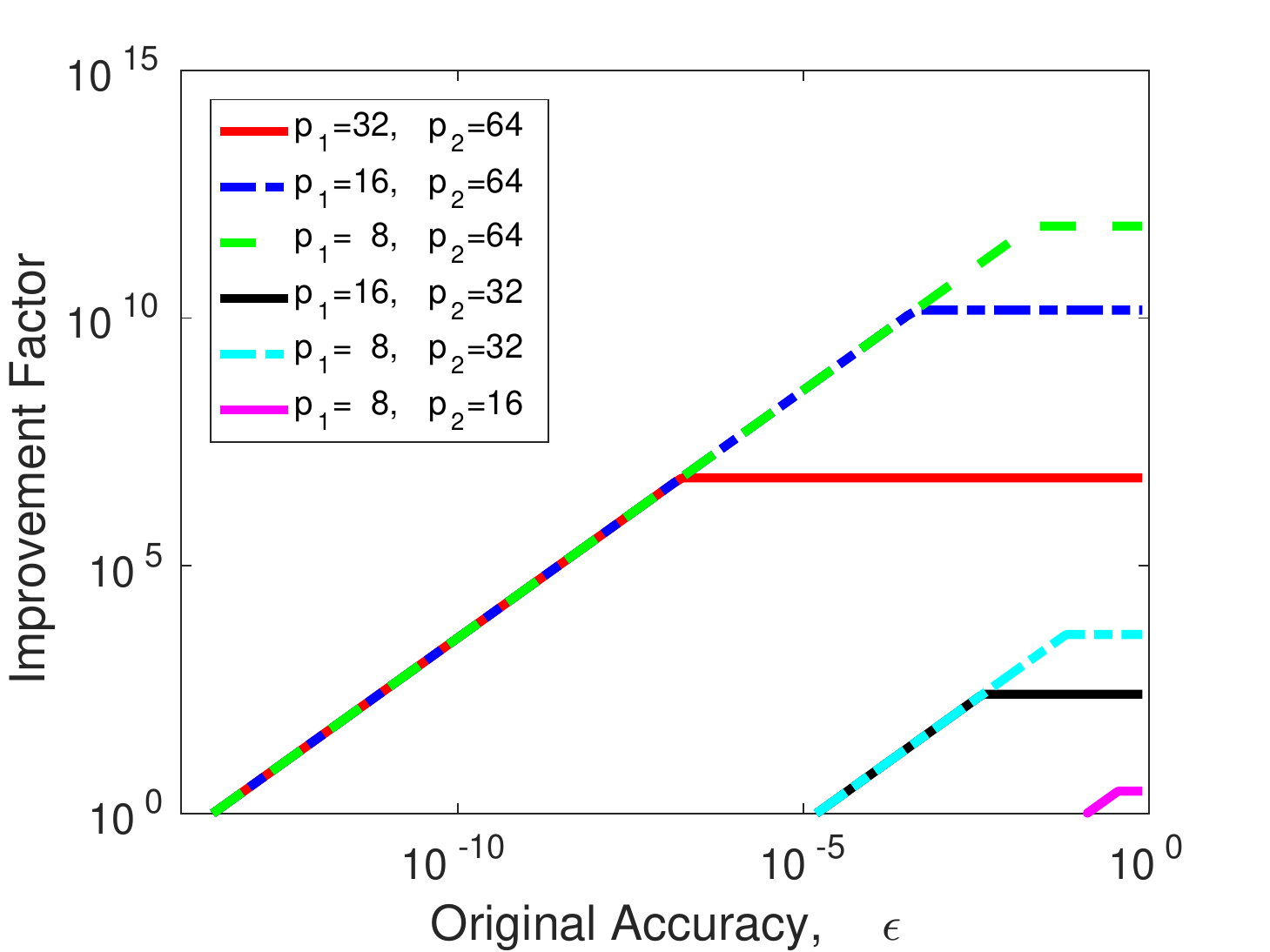}
\caption{Improvement factor bound \eqref{eq:impfact} for linear convergence 
rate as a function of $\epsilon$ for different hybrid precision levels (assumes 
$s_i=3\lg p_i+1$ and $\frac{E(p_1)}{E(p_2)}=\frac{p_1}{p_2}$).} 
\label{fig:impfactor-linear}
\end{figure}

Neglecting the reinvestment strategy for the moment, we also examine the 
energy expended for generic hybrid schemes. By using the optimal value 
$k_1=\frac{1}{\lg \lambda}\min\left\{p_1-s_1,\lg \frac{1}{\epsilon}\right\}$ 
and the corresponding
$$k_2=\frac{1}{\lg \lambda}\min\left\{p_2-s_2,
\max\left\{0,\lg \frac{1}{\epsilon}-k_1 \lg \lambda \right\}
\right\},$$
we can calculate via \eqref{eq:hybridp} the energy expended to attain an 
accuracy $\epsilon$.  This is illustrated in \Fig{fig:energy-acc-linear} 
for a variety of precision levels, under the assumptions $s_i=p_i/2$ and 
$E(p_i)\propto p_i$.

\begin{figure}[tb]
\centering\includegraphics[width=\linewidth]{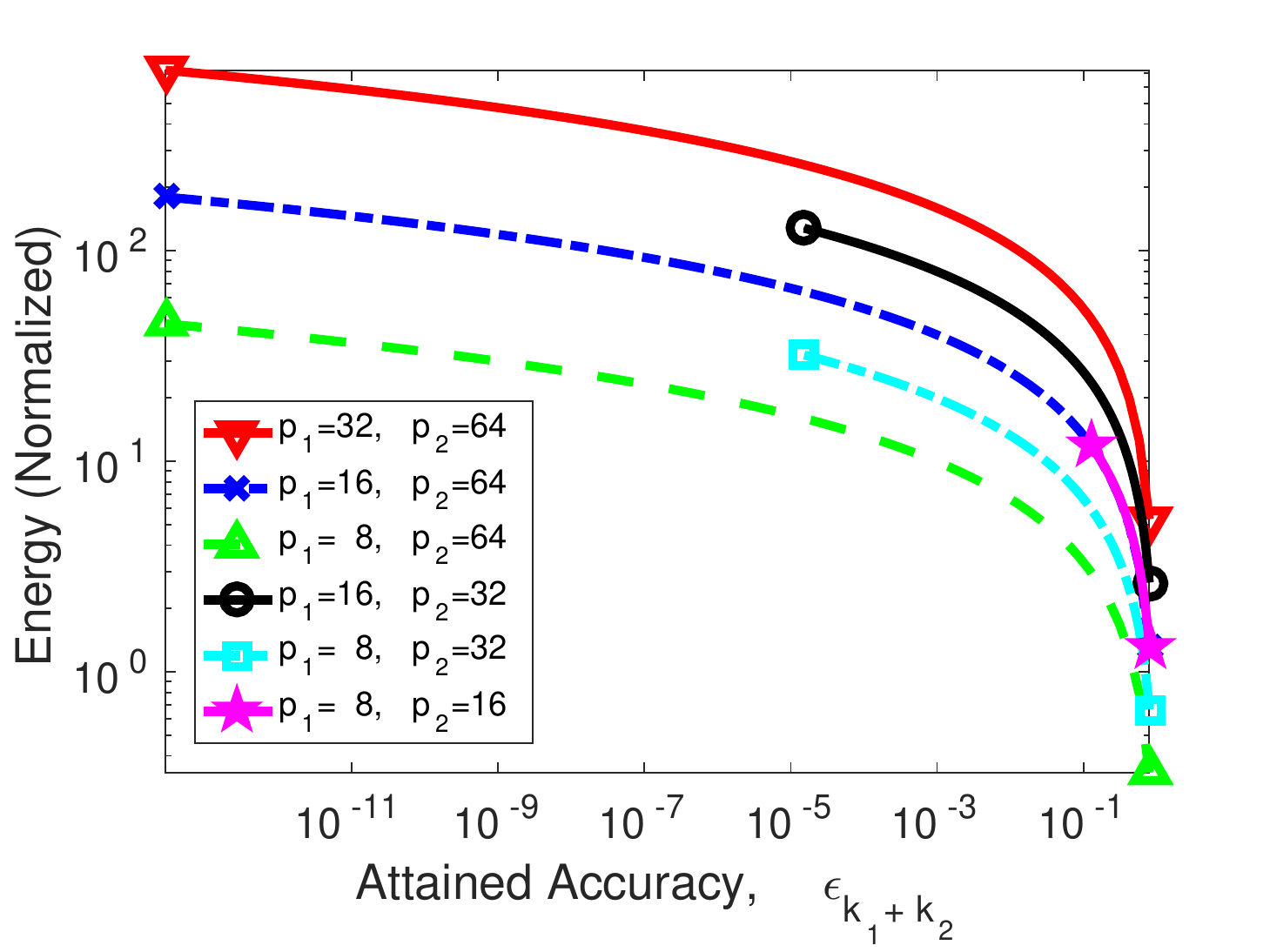}
\caption{Energy (less a normalizing constant) expended for linear 
convergence 
rate as a function of $\epsilon$ for different hybrid precision levels (assumes 
$s_i=3\lg p_i +1$ and $E(p)\propto p$).} 
\label{fig:energy-acc-linear}
\vspace{-0.2cm}
\end{figure}

\subsubsection*{Quadratic Convergence}
For an iterative method with a quadratic convergence rate, one has that 
$\epsilon_{k+1} \leq (\epsilon_k)^2/\lambda$, where $\lambda >1$. The
error after $k$ iterations is thus bounded by
$\epsilon_k \leq \lambda^{-2k+1}$. In order to satisfy $\epsilon_k \geq 
2^{-p+s}$, at most $\frac{p-s}{2\lg \lambda}+\frac{1}{2}$ iterations at a 
precision level $p$ are possible.  

Applying similar arguments as in the case of the linear 
convergence rate, we have that the accuracy of a 
hybrid, quadratically converging procedure is bounded by 
\begin{equation}
\begin{array}{rl}
\epsilon_{k_1+k_2}
&\leq 
\max\left\{ 2^{-p_2+s_2},
\lambda^{-\left(\frac{p_1-s_1}{\lg \lambda}+1\right)\left(1- 
\frac{E(p_1)}{E(p_2)}\right) - \frac{\lg 
\frac{1}{\epsilon}}{\lg \lambda}+1}
\right\} \\
&=
2^{-\min\left\{ p_2-s_2,
\left(p_1-s_1\right)\left(1- 
\frac{E(p_1)}{E(p_2)}\right) + \lg 
\frac{1}{\epsilon}-\frac{E(p_1)}{E(p_2)}\lg \lambda
\right\}}.
\end{array}
\label{eq:hybrid-acc2-q}
\end{equation}
Thus, either the higher-precision level's
attainable accuracy is binding, or the achieved accuracy is changed 
from that in \eqref{eq:hybrid-acc2}
by a 
modest factor $\lambda 2^{\frac{E(p_1)}{E(p_2)}}$. Therefore the 
improvement factor lower bound decreases by at most a factor $\lambda^{-1} 
2^{-\frac{E(p_1)}{E(p_2)}}$.

\section*{Conclusions and Outlook}

We have developed in this paper a somewhat paradoxical procedure for
reducing the errors in numerical computations by reducing the precision
of the floating point operations used. Our focus has been on using in 
the best possible manner a given energy budget. 
The paper illustrated one possible tradeoff between computation budget
and quality, namely the one achieved by changing numerical precision
of floating point numbers. As mentioned in the introduction, there are
many other potential ``knobs'' to trade off quality against
computation budget:  One can use different approximations in the
mathematical model, different  computation
methods, different discretizations of the continuous model, different levels of asynchrony,
etc. Each of these knobs has been studied in isolation. But there are
not independent; we are missing a methodology for finding the 
combination of choices for these knobs that achieve the best tradeoff
between quality and computation effort.

The ``dual'' problem, or reducing energy consumption for a given
result quality, is also important. Our results essentially show that
one can reduce energy consumption by a factor of $2.x$, without
affecting the quality of the result, by smarter use of single precision.
For decades, increased supercomputer performance has meant more double
precision floating point operations per second. This brute force approach is going to hit a
brick wall pretty soon. Smarter use of available computer resources is
going to be the main way of increasing the \emph{effective}
performance of supercomputers in the future.

Among the many scientific domains where effective supercomputing has come
to play a central role, none are perhaps more important than weather
prediction and climate modeling. Inexactness or phase I of our
approach has been shown in earlier work~\cite{palem-4, palem-date} to yield
benefits to weather prediction models with lower energy consumption
while preserving the quality of the prediction. This has spurred
interest among climate scientists who view inexactness through
precision reduction as a way of achieving speedups in the traditional
sense, and also cope with energy barriers~\cite{palmer,ECMWF}.
However, it is well understood that for serious advances in
model quality, weather and climate models need to be resolved at much
higher resolutions that is possible today with current computational
budgets including energy.  We hope that the new direction that the
results in this paper have demonstrated through the novel approach of
reinvestment to raising application quality significantly, will be a
harbinger for broader adoption of our two-phased approach by the
weather and climate modeling community. In particular, building on our
work here, the goal of selectively reducing precision resulting in
energy savings, while increasing the resolution of the weather and
climate models through energy reinvestment provides a path of
considerable societal value.

\section*{Acknowledgment}
This material is based upon work supported by 
the US Dept.\ of Energy, Office of Science, ASCR, under contract 
DE-AC02-06CH11357 and by DARPA Grant FA8750-16-2-0004. K. Palem's work was also supported in part by a Guggenheim fellowship.

\bibliographystyle{IEEEtran}
\bibliography{refs}

\end{document}